# The Effect of van der Waals interaction in Elastic Collision between Ps(1s) and H(1s)

Hasi Ray[1,2,3,*] , Anuradha De[3]  and Deparpita Roy[1]

[1]*Study Center, S-1/407, B. P. Township, Kolkata 700094, India*
[2]*Department of Physics, New Alipore College, Kolkata 700053, India*
[3]*Science Department, National Institute of TTT and Research, Kolkata 700106, India*

[*] hasi_ray@yahoo.com

**Abstract:** The modified static exchange model (MSEM) recently introduced by Ray [1] to study two-atomic collision processes at low and cold-energies, is used for detailed analysis of the long-range effects due to induced dipole-dipole van der Waals interaction between Ps and H atoms. The MSEM includes the non-adiabatic short-range effect due to electron-exchange and the long-range effect due to induced dynamic dipole polarisabilities of the atoms. The effective interatomic potential is highly sensitive to the minimum distance between the atoms ( $R_0$ ). The s-, p- and d- wave elastic phase shifts, corresponding partial cross sections, the scattering length and effective ranges are calculated and studied with the variation of the chosen least interatomic distance between them. It is found that the scattering length is highly sensitive to the effective interatomic potential that depends on the least interatomic distance. In addition the studies are made in search of Feshbach resonances. The observed interesting feature with the variation of $R_0$ in the triplet channel invites more accurate investigations if new physics.

# 1. Introduction

Very recently a modified static exchange model (MSEM) is introduced by Ray [1-2] to study the two-atomic collision processes at low and cold energies. The theory includes the non-adiabatic short-range effect due to electron-exchange and the long-range effect due to induced dipole polarisabilities of the atoms. The controlling of the s-wave scattering length in ultracold collision is an important problem to tune Feshbach resonances [3-12]. In ultracold system the kinetic energies of the atoms are negligibly small, so the interaction time is much longer than normal atomic interactions. If the density is ~ $10^{10}$ atoms/cm$^3$; the interatomic separation is ~ $10^{-4}$ to $10^{-3}$ cm, it is $10^4$ to $10^5$ times larger than atomic dimension ~ $10^{-8}$ cm [13]. As a result two of the slowly moving atoms can come close to each other when all others are far apart. In this approximation the atomic collision physics is able to provide reliable information about the cold atomic system [1-2]. The short-range effect due to electron exchange and the long range effect due to induced dynamic dipole polarisabilities of the atoms start to dominate as the system moves towards the colder energy region. In addition the need of quantum control of isolated atomic system with nanoscale localization instead of the collective motion of atoms is realized recently [14] as new possibilities for quantum control. The appropriate qubit generation for quantum computation [15] is again a challenging job today.

The well-known long-range interaction between two atoms is the van der Waals interaction. It is caused by the induced dipole polarisabilities of the atoms and is defined as

$$V_{van}(R) = -\frac{C_W}{R^6}, \qquad (1.1),$$

where $C_W$ is the *van der Waals coefficients* and $R$ is the interatomic distance [16]. The model form of van der Waals interaction used by Barker and Bransden [17] to study the quenching of ortho-Ps by He is :

$$V_{van}(R) = 0, \quad \text{if } R \langle R_0 ;$$
$$V_{van}(R) = -\frac{C_W}{R^6}, \quad \text{if } R \geq R_0, \quad \text{when } R_0 \to 0 \qquad (1.2).$$

So the minimum value of interatomic distance ($R_0$) is an important parameter to determine the strength of effective interatomic potential. **Figure 1** describes the typical variation of the effective interatomic potential as a function of interatomic distance ($R$). When two atoms are far apart i.e. $R \to \infty$ the effective interatomic potential is almost zero indicating almost no interaction between the atoms. When they proceed to each other i.e. $R$ decreases, the effective interatomic potential starts to be more and more negative and reaches a minimum value at $R = R_{min}$, i.e. maximum attraction between the atoms. The attraction gradually decreases as $R \langle R_{min}$, so the potential gradually increases, becomes zero when the interatomic distance is $R_0$. The interatomic potential starts to be sharply positive as $R \langle R_0$ due to strong static Coulomb interaction between the atoms. So the atoms begin repelling each other strongly and can not proceed further towards themselves. As a result, the minimum value of the interatomic distance is $R_0$. The similar concept is discussed in Lennard-Jones 6-12 potential [18]. It is defined as

$$V_{LJ}(R) = -4\varepsilon \left\{ \left(\frac{\sigma}{R}\right)^6 - \left(\frac{\sigma}{R}\right)^{12} \right\} \qquad (1.3),$$

when $\varepsilon$ and $\sigma$ are *the Lennard-Jones parameters*. In the above expression, the first term corresponds to the long-range van der Waals interaction part and the second term corresponds to the short-range electron-electron exchange part and it is repulsive. In principle the short range electron-electron exchange force is repulsive if the electron spins form a triplet (-) state and it is attractive if they form a singlet (+) state.

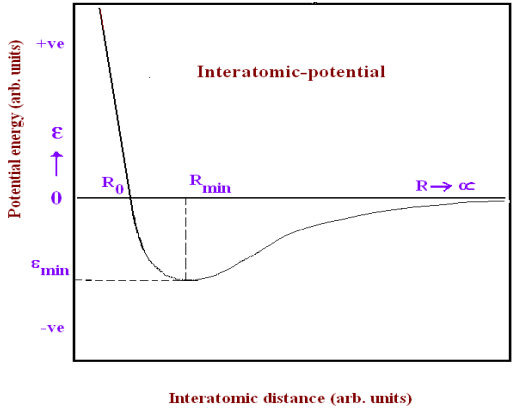

**Figure 1.** The variation of interatomic potential with interatomic distance R.

The present paper deals the elastic collision between Ps and H when both are in ground states. In the coupled-channel methodology, the eigen-state expansion to define the system wavefunction and an approach like the Hartree-Fock variational method to project out different channels are used to get the set of coupled integro-differential equations. In differential approach one uses iterative method and partial wave analysis to find the unknown coefficients defined in eigen-state expansion. However the number of coupled equations is restricted there by the number of bound states [19] taken into account. In integral approach one uses the help of Lippman-Schwinger equation to get the coupled integral equations [19]. Either the configuration space [20] or the momentum space [1-4, 19, 21-23] can be used to form the equations. However in the momentum space formalism [19] the convergence problem is much easier to overcome than the coordinate space formalism. When the exchange amplitude is combined (adding or subtracting) with the direct first-Born amplitude, it is called the Born-Oppenhimer (BO) amplitude following the nomenclature of Ray and Ghosh [21] and accordingly the singlet (+) and triplet (-) channels are defined. The BO ($\pm$) amplitudes act as input in the coupled-channel methodology to obtain the desired unknown amplitudes for the singlet (+) and triplet (-) channels respectively. The partial wave analysis and angular momentum algebra are used to reduce the three-dimensional coupled integral equation into the one-dimensional form. The partial wave contributions: L=0 is defined as the s-wave, L=1 as the p-wave and so on. The effective range theory [5] is useful to derive the scattering length ($a$) and the effective range ($r_0$) utilizing the variation of the s-wave elastic phase-shift ($\delta_0$) with the incident energy. The s-, p-, d- wave elastic phase shifts, the corresponding s-, p-, d- wave elastic cross sections, the scattering lengths and effective ranges for both the singlet and triplet channels are evaluated choosing different values of $R_0$.

## 2. Theory

In the static-approximation (SM) one solve the Schrodinger equation

$$H\psi = E\psi \qquad (2.1)$$

using the eigen state expansion to write the system wavefunction as

$$\psi = \sum_i a_i \phi_i \qquad (2.2)$$

when $a_i$ is the unknown coefficient and $\varphi_i$ is the channel wavefunction that considers only the direct elastic channel. The elastic channel is defined so that $|\vec{k}_i| = |\vec{k}_f|$ if $|\vec{k}_i|$ and $|\vec{k}_f|$ are the initial and final momenta of the projectile.

In **Figure 2**, the proton $p^+$ of H and the positron $e^+$ of Ps are placed at points A and B respectively; e1 and e2 are the two electrons ($e^-$) attached to the nuclei A and B which are at a distance $r_{1A}$ and $r_{2B}$ from A and B respectively. Accordingly $\vec{r}_{2A}$ and $\vec{r}_{1B}$ are defined. $\vec{R}$ is the vector joining A and B and $\vec{R}'$ is the vector joining the center of masses of the two atoms so that

$$\vec{R}' = \vec{R} + \frac{m_e}{m_A + m_e}\vec{r}_{1A} - \frac{m_e}{m_B + m_e}\vec{r}_{2B} \tag{2.3}$$

when $m_A$, $m_B$ and $m_e$ are the masses of $p^+$, $e^+$ and $e^-$ respectively.

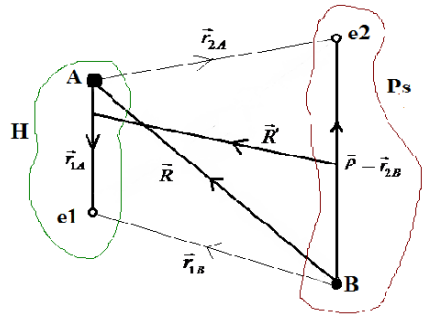

**Fig. 2.** The picture of a Ps-H system.

The modifying amplitude due to long-range van der Waals interaction that is added to the BO amplitude [21] to modify the static-exchange model (SEM) potential in the present MSEM [1] is defined as

$$B_{van} = \int d\vec{r}_{1A}\int d\vec{r}_{2B}\int d\hat{R}\int_{R=R_0}^{\infty} dR.R^2\left[\psi^*(\vec{R},\vec{r}_{1A},\vec{r}_{2B})\{-\frac{C_W}{R^6}\}\psi(\vec{R},\vec{r}_{1A},\vec{r}_{2B})\right] \tag{2.4}$$

if $\psi$ represents the system wavefunction.. The long-range modifying amplitude defined above depends on $R_0$. How the value of $R_0$ should be determined is an important question. There is no good literature. However in the cold-atomic system as the atoms are very slow, the density of the atoms in the system seems to control the value of $R_0$. Different values of $R_0$ starting from $2.2\,a_0$ to $10\,a_0$ are used to vary the strength of long-range potential; the symbol $a_0$ represents the Bohr radius.

The SEM includes the non-adiabatic short-range effects due to exchange or antisymmetry of the system electrons. The initial and final channel wave functions are defined as

$$\psi_i = e^{i\vec{k}_i \cdot \vec{R}}\phi_{1s}(r_{1A})\eta_{1s}(r_{2B}) \tag{2.5}$$

$$\psi_f = (1 \pm P_{12})e^{i\vec{k}_f \cdot \vec{R}}\phi_{1s}(r_{1A})\eta_{1s}(r_{2B}) \tag{2.6}$$

Here $\phi_{1s}(r_{1A})$ and $\eta_{1s}(r_{2B})$ are the ground state wave functions of H and Ps respectively and $P_{12}$ is the exchange ( or antisymmetry) operator. In the present concept of scattering theory, a free projectile coming from infinity e.g. a plane wave when enters into the Coulomb field of target, it suffers collision and accumulate information as phase shift. So only the final state wavefunction needs to be antisymmetrized. To define the SEM potential [21], we include all the four Coulomb interaction terms: the interaction between proton and positron, the interaction between proton and electron 2, the interaction between positron and electron 1, and the interaction between electron 1 and electron 2 in the direct channel and accordingly in the rearrangement channel exactly. We use atomic masses. The discussion about the importance of atomic and nuclear masses are available in the literature [24].

The formally exact Lippman-Schwinger type coupled integral equation for the scattering amplitude in momentum space [19] is given by :

$$f^{\pm}_{n'1s,n1s}(\vec{k}_f,\vec{k}_i) = B^{\pm}_{n'1s,n1s}(\vec{k}_f,\vec{k}_i) - \frac{1}{2\pi^2}\sum_{n''}\int d\vec{k}'' \frac{B^{\pm}_{n'1s,n''1s}(\vec{k}_f,\vec{k}'')f^{\pm}_{n''1s,n1s}(\vec{k}'',\vec{k}_i)}{\vec{k}^2_{n''1s}-\vec{k}''^2+i\varepsilon} \qquad (2.7).$$

Here $B^{\pm}$ are the well known Born-Oppenheimer (BO) scattering amplitude in the singlet (+) and triplet (-) channels respectively. Similarly $f^{\pm}$ indicate the unknown SEM scattering amplitudes for the singlet and triplet channels. The BO amplitude is defined as

$$B^{\pm}_{n'1s,n1s}(\vec{k}_f,\vec{k}_i) = -\frac{\mu}{2\pi}\int d\vec{R}d\vec{r}_1 d\vec{r}_2 \psi^*_f(\vec{R},\vec{r}_1,\vec{r}_2) V(\vec{R},\vec{r}_1,\vec{r}_2)\psi_i(\vec{R},\vec{r}_1,\vec{r}_2) \qquad (2.8),$$

when $\mu$ is the reduced mass of the system. The Coulomb interaction between the atoms in the direct and rearrangement channels are expressed as

$$V_{Direct}(\vec{R},\vec{r}_{1A},\vec{r}_{2B}) = \frac{p^+e^+}{R} - \frac{p^+e^-_2}{|\vec{R}-\vec{r}_{2B}|} - \frac{e^+e^-_1}{|\vec{R}+\vec{r}_{1A}|} + \frac{e^-_1 e^-_2}{|\vec{R}+\vec{r}_{1A}-\vec{r}_{2B}|} \qquad (2.9),$$

$$V_{Exchange}(\vec{R},\vec{r}_{1A},\vec{r}_{2B}) = \frac{p^+e^+}{R} - \frac{p^+e^-_1}{|\vec{r}_{1A}|} - \frac{e^+e^-_2}{|\vec{r}_{2B}|} + \frac{e^-_1 e^-_2}{|\vec{R}+\vec{r}_{1A}-\vec{r}_{2B}|} \qquad (2.10),$$

respectively. Here the magnitudes of all the Coulomb terms in the numerators in equation (2.9) and (2.10) are equal to unity in atomic unit (a.u.). The s-wave elastic phase shift and the corresponding cross section are studied in the energy region starting from $1\times10^{-4}$ eV to 0.1 eV for both the singlet and triplet states of system electrons. The most accurate value of $C_W = 34.785$ a.u. reported by Mitroy and Bromley [25] is adapted.

The effective range theory expresses s-wave elastic phase shift as a function of scattering length and projectile energy so that

$$k\cot[\delta_0(k)] = -\frac{1}{a} + \frac{1}{2}r_0 k^2 + .... \qquad (2.11)$$

where $\delta_0$ is the s-wave elastic phase shift, $\vec{k}$ is the incident momentum, $a$ is the scattering length and $r_0$ is the range of the potential. Accordingly, the scattering length is defined as

$$a = -\lim_{k\to 0}\left\{\frac{\tan[\delta_0(k)]}{k}\right\} \qquad (2.12)$$

The scattering length, $a \langle 0$ indicates no possibility of binding in the system. The positive scattering length i.e. $a \rangle 0$ indicates a possibility of the presence of a Feshbach resonance and binding. A rapid change in phase-shift by $\pi$ radian is an indication of the presence of a Feshbach resonance [3,5,6].

## 3. Results and discussion

We reproduce the SEM data of Ray et al [21] using the present MSEM code and switching-off the van der Waals interaction term, for the values of $k$ equal to .1, .2, .3, .4, .5, .6, .7, .8 a.u. We use the MSEM code switching-on the van der Waals interaction and varying the interatomic distance $R_0$=10 a.u., 7 a.u., 5 a.u., 4 a.u., 3 a.u. and 2.5 a.u. respectively to calculate the s-wave elastic phase shifts and corresponding partial cross sections. A comparison is made of the MSEM data with the SEM data for the s-wave elastic singlet (+) phase shifts in **Table 1(a)** and triplet phase shifts (-) in **Table 1(b)**. The data are showing a very good agreement with the existing physics [26], all the phase shift results are gradually increasing with the increase of strength of van der Waals interaction as $R_0$ decreases. In **Figure 3**, the s-wave elastic phase shifts for both the singlet and triplet channels using present MSEM theory are compared with the SEM data at too low energy region. The corresponding s-wave elastic cross sections are presented in **Figure 4a** for the singlet channel and in **Figure 4b** for the triplet channel.

**Table 1(a)** Comparison of the SEM data of the s-wave elastic singlet phase shift ($\delta_0^+$) in radian with the MSEM data varying the inter- atomic separation ($R_0$).

| $k$ (a.u.) | SEM | MSEM with $R_0$=10 a.u. | MSEM with $R_0$=7 a.u. | MSEM with $R_0$=5 a.u. | MSEM with $R_0$=4 a.u. | MSEM with $R_0$=3 a.u. | MSEM with $R_0$=2.5 a.u. |
|---|---|---|---|---|---|---|---|
| 0.1 | 2.457 | 2.458 | 2.458 | 2.459 | 2.467 | 2.521 | 2.604 |
| 0.2 | 1.927 | 1.928 | 1.929 | 1.930 | 1.938 | 2.005 | 2.118 |
| 0.3 | 1.539 | 1.540 | 1.542 | 1.543 | 1.547 | 1.604 | 1.717 |
| 0.4 | 1.239 | 1.240 | 1.243 | 1.245 | 1.247 | 1.286 | 1.390 |
| 0.5 | 0.998 | 0.998 | 1.000 | 1.006 | 1.007 | 1.031 | 1.116 |
| 0.6 | 0.798 | 0.798 | 0.799 | 0.808 | 0.811 | 0.823 | 0.890 |
| 0.7 | 0.631 | 0.631 | 0.632 | 0.641 | 0.649 | 0.653 | 0.700 |
| 0.8 | 0.491 | 0.491 | 0.492 | 0.499 | 0.511 | 0.516 | 0.545 |

**Table 1(b)** Comparison of the SEM data of the s-wave elastic triplet phase shift ($\delta_0^-$) in radian with the MSEM data varying the inter- atomic separation ($R_0$).

| $k$ a.u. | SEM | MSEM with $R_0$=10 a.u. | MSEM with $R_0$=7 a.u. | MSEM with $R_0$=5 a.u. | MSEM with $R_0$=4 a.u. | MSEM with $R_0$=3 a.u. | MSEM with $R_0$=2.5 a.u. |
|---|---|---|---|---|---|---|---|
| 0.1 | -0.247 | -0.245 | -0.241 | -0.234 | -0.225 | -0.210 | -0.194 |
| 0.2 | -0.489 | -0.488 | -0.482 | -0.469 | -0.453 | -0.421 | -0.390 |
| 0.3 | -0.721 | -0.721 | -0.717 | -0.700 | -0.677 | -0.630 | -0.580 |
| 0.4 | -0.940 | -0.940 | -0.938 | -0.921 | -0.894 | -0.831 | -0.761 |
| 0.5 | -1.143 | -1.143 | -1.142 | -1.129 | -1.098 | -1.020 | -0.929 |
| 0.6 | -1.330 | -1.329 | -1.329 | -1.320 | -1.289 | -1.197 | -1.081 |
| 0.7 | -1.499 | -1.499 | -1.498 | -1.494 | -1.466 | -1.362 | -1.222 |
| 0.8 | -1.653 | -1.652 | -1.651 | -1.649 | -1.627 | -1.516 | -1.351 |

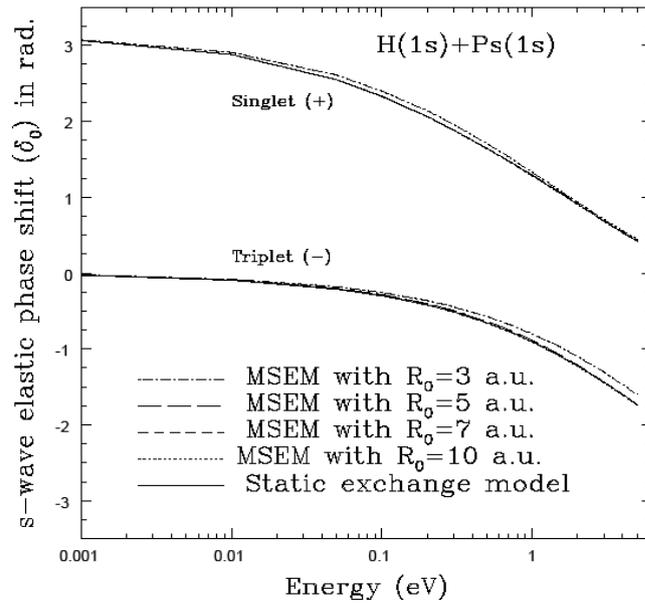

**Figure 3.** The comparison of present s-wave elastic phase-shift using the SEM and the MSEM for different values of $R_0$.

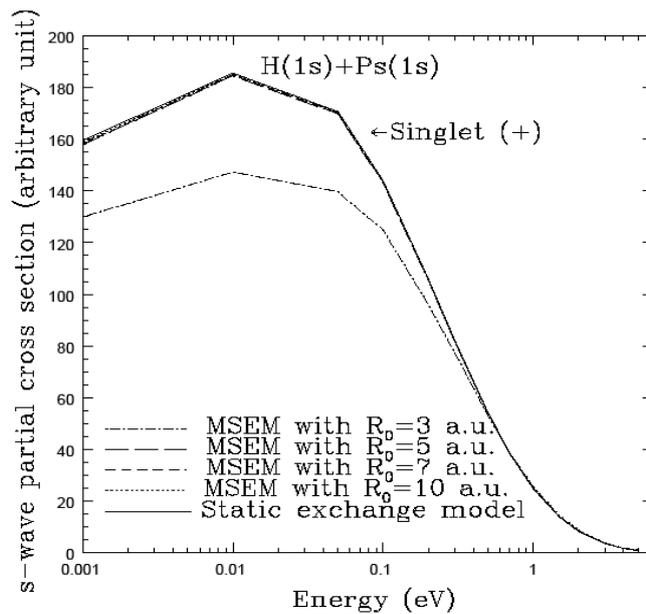

**Figure 4a.** The s-wave partial cross section in arbitrary unit for singlet state of system electrons.

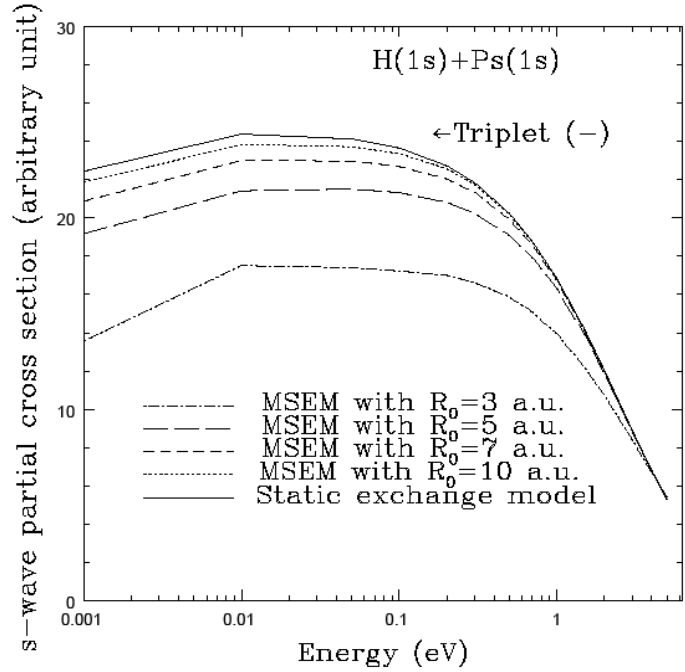

**Figure 4b.** The s-wave partial cross section in arbitrary unit for triplet state of system electrons.

To evaluate the scattering lengths the effective range theory is useful. The $k\cot\delta_0$ is plotted against $k^2$ to evaluate $-1/a$ in **Figure 5a** for singlet channel and in **Figure 5b** for triplet channel. The scattering length ($a$) vary systematically with varying $R_0$ chosen as $3a_0$, $5a_0$, $7a_0$ and $10a_0$. The variation of the computed scattering lengths and ranges with the interatomic potential using different values of $R_0$ are presented in **Table 2** and compared with the SEM data and available data [27-32].

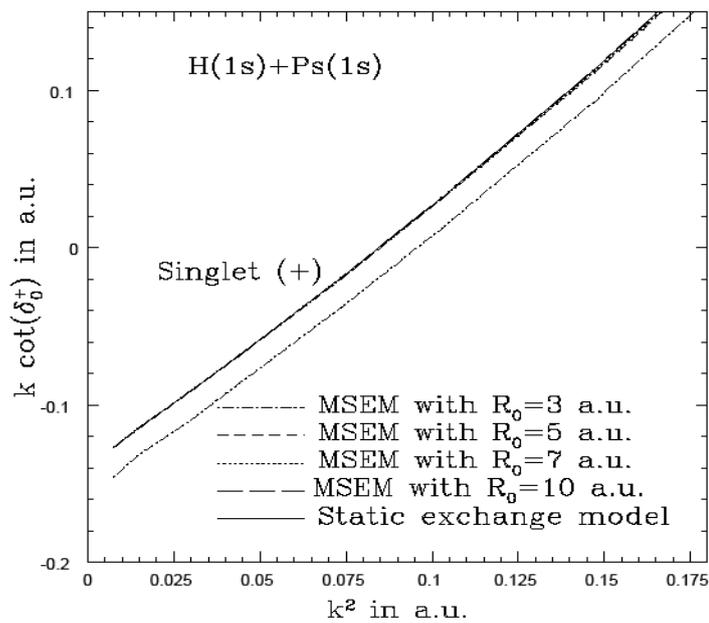

**Figure 5a.** The $k\cot\delta_0$ versus $k^2$ plot in atomic units for singlet channel.

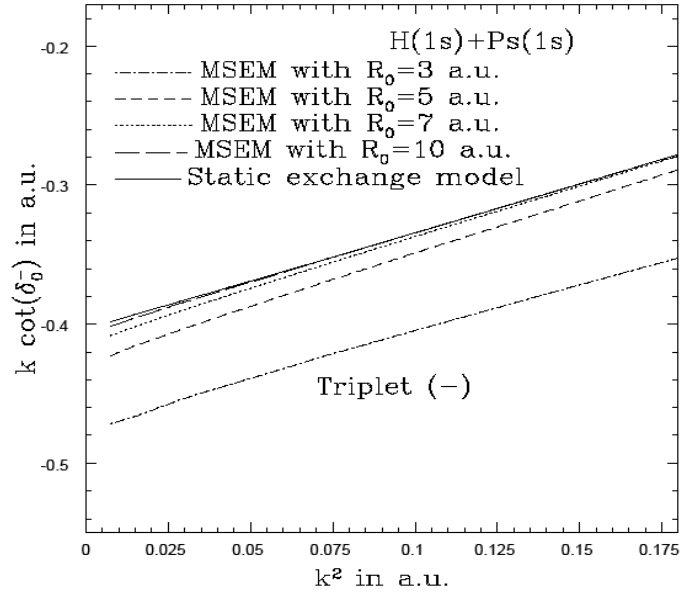

**Figure 5b.** The $k\cot\delta_0$ versus $k^2$ plot in atomic units for triplet channel.

When $R_0 = 10a_0$ or greater, all the data almost coincide with the SEM data. There are always a very small difference between the SEM data and MSEM data almost upto $R_0 = 50a_0$ indicating the long-range behaviour of the potential. Here all the scattering lengths are positive. The variation of triplet scattering length is more sensitive to long-range interaction than singlet scattering length [26]. A very large number of mesh points are required to study the resonances.

**Table 2.** The s-wave elastic scattering lengths ($a$) and effective ranges ($r_0$) in atomic units for singlet (+) and triplet (−) spin-configurations of system electrons.

| Scattering parameters | SEM data (a.u.) | MSEM data when $R_0 = 10a_0$ (a.u.) | MSEM data when $R_0=7a_0$ (a.u.) | MSEM data when $R_0=5a_0$ (a.u.) | MSEM data when $R_0 = 4a_0$ (a.u.) | MSEM data when $R_0 = 3a_0$ (a.u.) | MSEM data when $R_0 = 2.5a_0$ (a.u.) | Data of others (a.u.) |
|---|---|---|---|---|---|---|---|---|
| $a$ (+) | 7.25 | 7.22 | 7.19 | 7.14 | 7.04 | 6.17 | 5.32 | 4.5[a], 5.22[b], 5.20[c], 5.84[d], 3.49[e], 4.30[f] |
| $r_0$ (+) | 3.79 | 3.78 | 3.77 | 3.75 | 3.74 | 3.73 | 3.48 | 2.2[a], 2.90[d], 2.08[f] |
| $a$ (−) | 2.49 | 2.45 | 2.43 | 2.36 | 2.27 | 2.12 | 2.02 | 2.36[a], 2.41[b], 2.45[c], 2.32[d], 2.46[e], 2.20[f] |
| $r_0$ (−) | 1.42 | 1.41 | 1.40 | 1.38 | 1.34 | 1.28 | 1.19 | 1.31[a] |

[a] Stabilization calculation of Drachman and Houston [27].
[b] Close-coupling calculation of Sinha, Basu and Ghosh [28].
[c] R-matrix calculation of Blackwood, McAlinden, and Walters [29].
[d] Kohn variational calculation of Page [30].
[e] Variational calculation of Adhikari and Mandal [31].
[f] Stochastic variational calculation of Ivanov et al [32].

In addition, we study the Feshbach resonances at very low energy region $10^{-4}$ to $10^{-2}$ for triplet channels varying the values of $R_0$ and using a very fine mesh-points. Very interesting features are observed. The s-wave elastic phase shifts are displayed in figure 6(a) using static exchange model (SEM), 6(b) using present MSEM with $R_0$ =6 a.u., 6(c) using present MSEM with $R_0$ =5 a.u., 6(d) using present MSEM with $R_0$ =4 a.u., 6(e) using present MSEM with $R_0$ =3 a.u., 6(f) using present MSEM with $R_0$ =2.5 a.u., 6(g) using present MSEM with $R_0$ =2.3 a.u., 6(h) using present MSEM with $R_0$ =2.2 a.u. respectively. The corresponding cross sections appear in Figures 7(a), 7(b), 7(c), 7(d), 7(e), 7(f), 7(g) and 7(h). An interesting behaviour both in phase shifts and corresponding partial cross sections at the energy region very close to E = $1.25 \times 10^{-4}$ eV continues its presence when $R_0$ varies from 5 to 2.5 a.u. There is a dip in cross section when $R_0$ =5 a.u., but that dip transforms into a peak when $R_0$ is less than 5 a.u. When $R_0$ =2.5 a.u. the cross section curve shows almost similar behaviour with SEM data. The parameter $R_0$ appears to be the determining factor to occur such behaviour in the system.

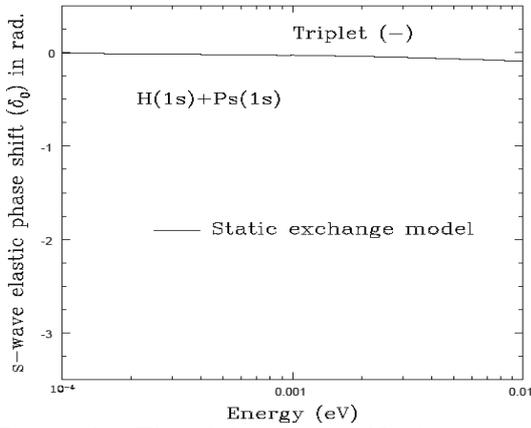

**Figure 6a.** The triplet phase shifts (s-wave) using the SEM.

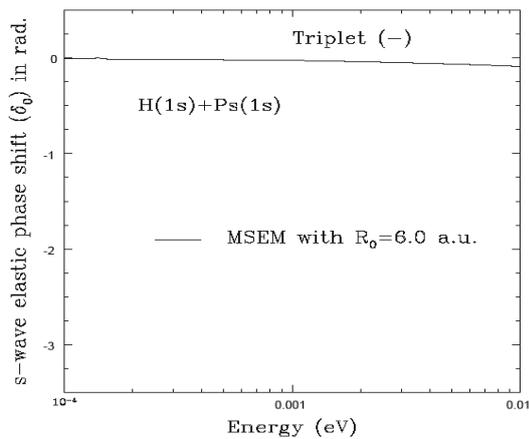

**Figure 6b.** The triplet phase shifts (s-wave) using the MSEM with $R_0$ = 6 a.u.

.

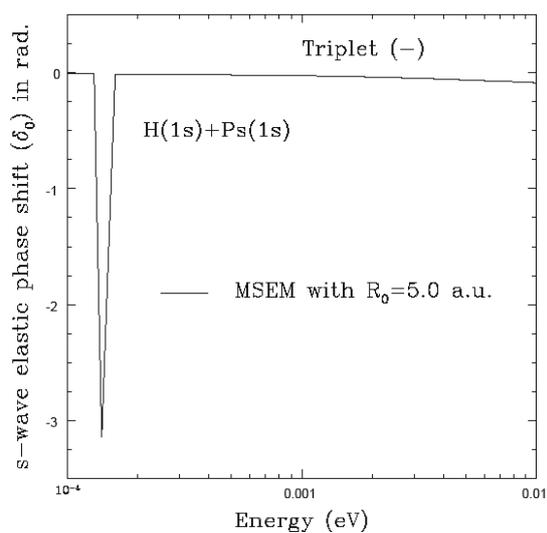
**Figure 6c.** The triplet phase shifts (s-wave) using the MSEM with $R_0 = 5$ a.u.

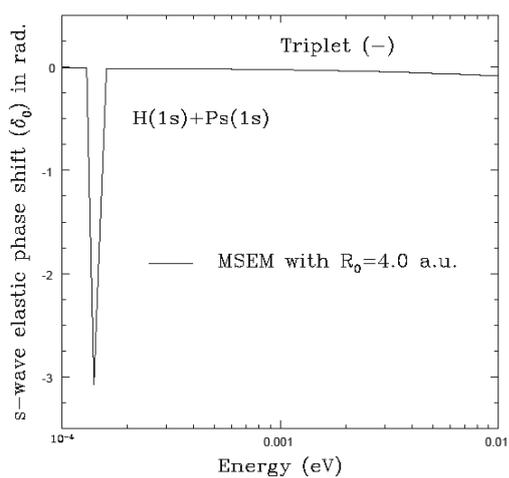
**Figure 6d.** The triplet phase shifts (s-wave) using the MSEM with $R_0 = 4$ a.u.

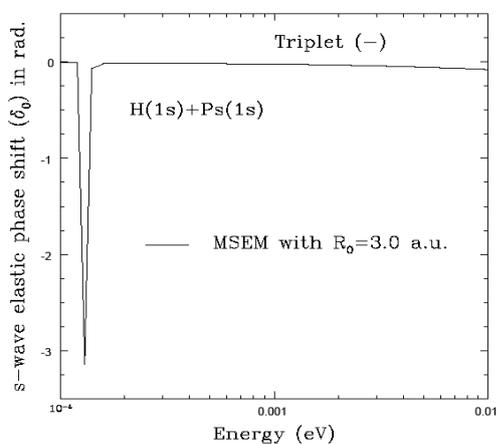
**Figure 6e.** The triplet phase shifts (s-wave) using the MSEM with $R_0 = 3$ a.u.

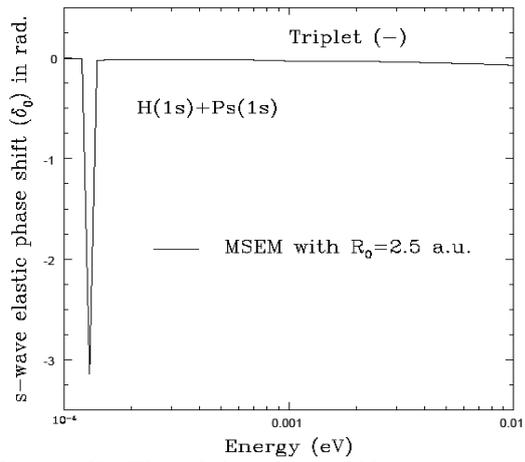

**Figure 6f.** The triplet phase shifts (s-wave) using the MSEM with $R_0 = 2.5$ a.u.

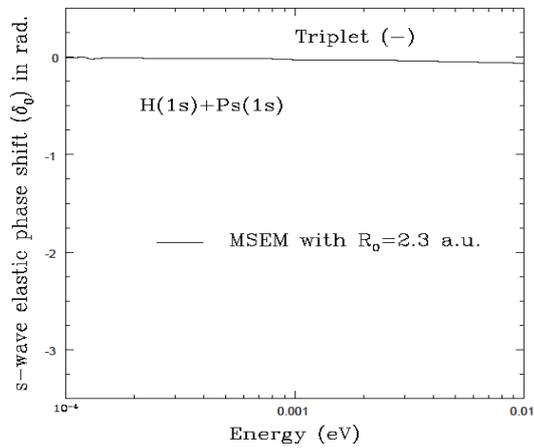

**Figure 6g.** The triplet phase shifts (s-wave) using the MSEM with $R_0 = 2.3$ a.u.

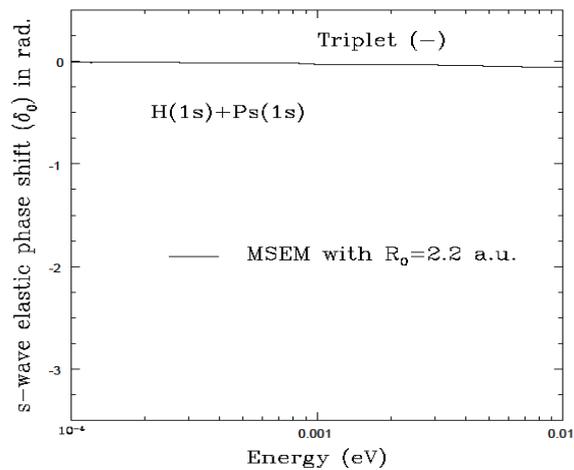

**Figure 6h.** The triplet phase shifts (s-wave) using the MSEM with $R_0 = 2.2$ a.u.

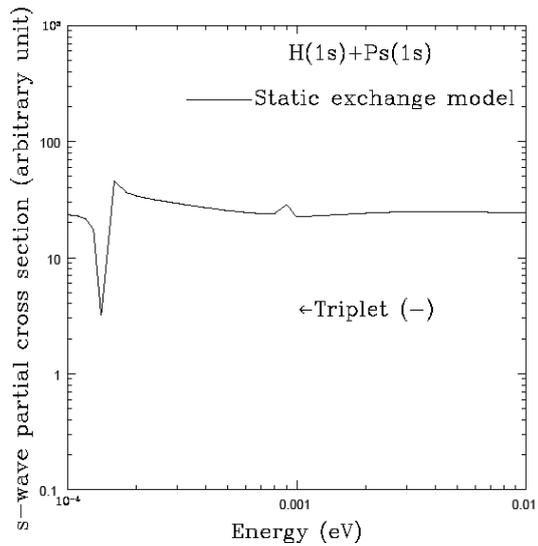

**Figure 7a.** The triplet cross sections (s-wave) using the SEM.

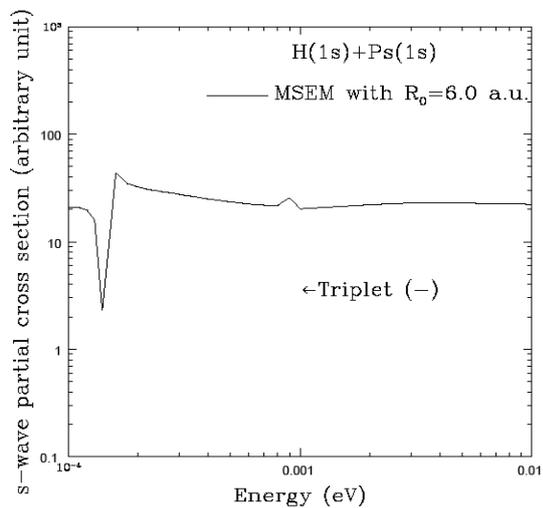

**Figure 7b.** The triplet cross sections (s-wave) using the MSEM with $R_0 = 6$ a.u.

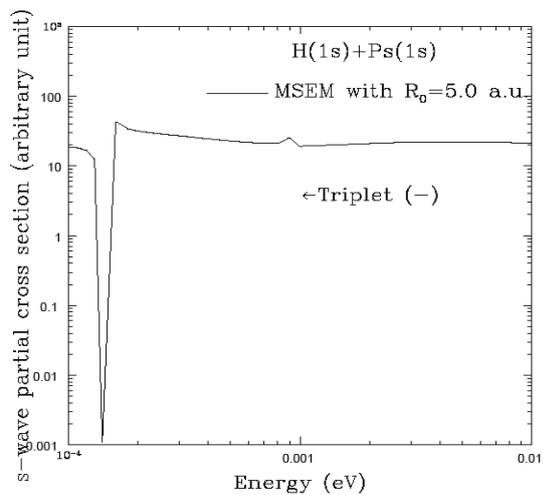

**Figure 7c.** The triplet cross sections (s-wave) using the MSEM with $R_0 = 5$ a.u.

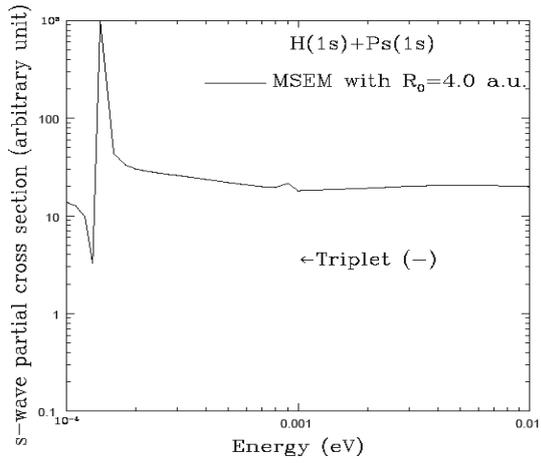
**Figure 7d.** The triplet cross sections (s-wave) using the MSEM with $R_0 = 4$ a.u.

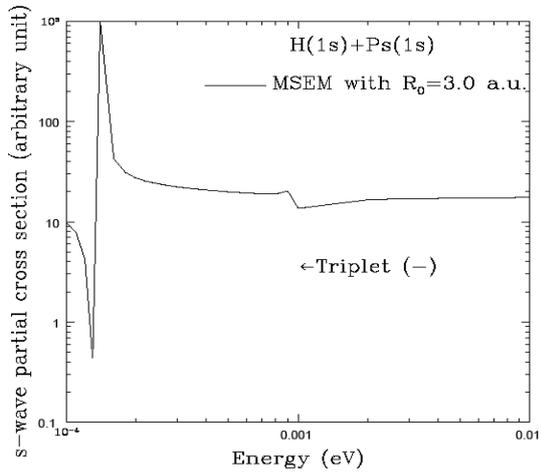
**Figure 7e.** The triplet cross sections (s-wave) using the MSEM with $R_0 = 3$ a.u.

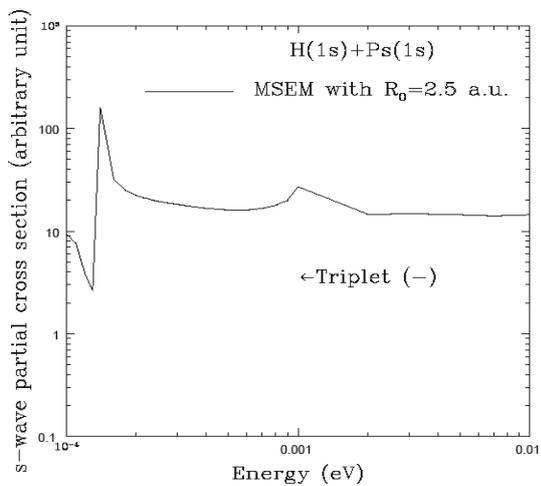
**Figure 7f.** The triplet cross sections (s-wave) using the MSEM with $R_0 = 2.5$ a.u.

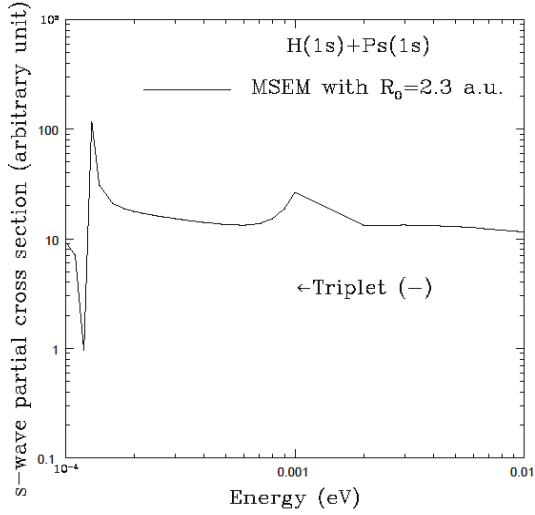

**Figure 7g.** The triplet cross sections (s-wave) using the MSEM with $R_0 = 2.3$ a.u.

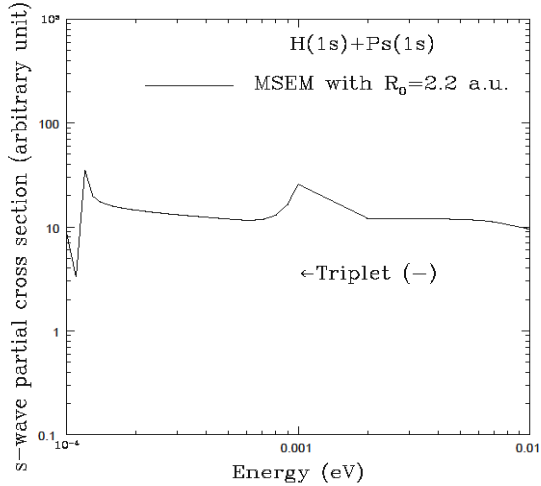

**Figure 7h.** The triplet cross sections (s-wave) using the MSEM with $R_0 = 2.2$ a.u.

## 4. Conclusions

A modified static exchange model (MSEM) is used for detailed analysis of the long-range potential due to van der Waals interaction between atoms at low/cold energis. The model is applied on positronium (Ps) and hydrogen (H) elastic collision when both the atoms are in ground states. Since both the atoms are hydrogen like and one of them Ps has light mass and high polarizability, the present system is highly useful to extract the basic new physics. We study the s-wave elastic phase shift and the corresponding cross section in the energy range $10^{-4}$ eV to 0.1 eV using the present model (MSEM) and the static exchange model (SEM). The variation of scattering length and range are studied with the variation of $R_0$ and compared with SEM and other available data [27-32]. The scattering lengths are found to be highly sensitive to the effective interatomic potential that varies with $R_0$. Again the parameter $R_0$ is responsible to occur an interesting behaviour close to energ E = $1.25 \times 10^{-4}$ eV. Such kind of behaviour in the triplet channel in Ps-H system is interesting to search new physics. So more accurate investigations are invited.

**Acknowledgments**

HR acknowledges the support of Department of Science and Technology (DST) India  Grant no. SR/WOSA/PS-13/2009.